\documentclass{aa}
\usepackage{lineno}     
\usepackage{linenoaa}   
\usepackage{graphicx}
\usepackage{mathrsfs}
\usepackage{txfonts}
\usepackage{subcaption}         
\usepackage{float} 
                                
\usepackage{lscape}             
                                
\usepackage{placeins}

\usepackage{wasysym}

\titlerunning{Devolatilization trends in Rocky Exoplanets}
\authorrunning{Reinarz et al.}

\begin{document}

   \title{The compositional link between rocky exoplanets and their host stars is consistent with Earth-like devolatilization}

\author{Yared Reinarz\inst{1,2}
        \and Lorena Acu\~na-Aguirre\inst{1}
        \and Haiyang S. Wang\inst{3}
        }
\institute{Max-Planck-Institut f\"ur Astronomie, K\"onigstuhl 17, Heidelberg 69117, Germany\\
             \email{reinarz@mpia.de; acuna@mpia.de}
           \and Heidelberg University, Grabengasse 1, 69117 Heidelberg, Germany
           \and Center for Star and Planet Formation, Globe Institute, University of Copenhagen, \O ster Voldgade 5-7, 1350 Copenhagen, Denmark\\
             \email{haiyang.wang@sund.ku.dk}
        }
   \date{Accepted October 30, 2026}

\abstract
{
Previous work found a higher bulk iron-to-silicon (Fe/Si) ratio in rocky exoplanets relative to their host star, traditionally explained by iron enrichment. An alternative explanation is devolatilization of the planetary abundance ratios with respect to the host star, where volatile elements such as oxygen and sulfur are progressively depleted compared to more refractory elements such as iron and magnesium, as in Earth. In this work, we explore whether these elemental molar ratios are consistent with a devolatilization trend. We revisit the estimation of the core mass fraction of 14 rocky exoplanets with masses between 0.5 and 10 Earth masses, focusing on planets where stellar abundances of Fe, Si, Mg, O, and S are available from stellar spectroscopy in the literature. We use interior structure retrievals with two independent forward models, the Marseille Super-Earth Interior model and pyExoInt, to estimate compositional parameters, including elemental molar ratios, from mass and radius measurements. We find that 8 out of 14 exoplanets in our sample present an Earth-like devolatilization trend, while only one planet shows a 1:1 relationship between the planetary and the stellar abundances. This suggests that devolatilization offers a plausible explanation for the observed non-1:1 relation. The two interior models are in good agreement on the devolatilization trend, demonstrating that our derived trends are robust. The main limiting factor in constraining refractory compositions between rocky planets and their host star is the small sample with constrained host stellar abundances and the lack of variety in orbital distances. The \textit{PLATO} mission will help solve this problem by extending the sample of characterized rocky planets to orbital periods of up to one year. The variety in refractory elemental ratio trends may be indicative of a wide diversity in bulk composition, outgassing and other mantle properties relevant for habitability.
}


   \maketitle

\nolinenumbers

\section{Introduction}

Since the first exoplanet discovery around a main-sequence star \citep{1995Natur.378..355M}, over 6000 exoplanets have been confirmed through various detection techniques \citep{Fischer_2014,Christiansen_2025}. To date, transit photometry \citep{Charbonneau_2000, Borucki_2010, Ricker_2014} and radial-velocity spectroscopy \citep{1995Natur.378..355M, Wright_2018} are the two most successful methods for detecting exoplanets. Their combination yields planetary masses, radii, and orbital parameters, from which the bulk density and equilibrium temperature can be obtained. These methods have intrinsic observational biases, however: larger planets orbiting smaller stars at close distances are more easily detected than small planets orbiting larger stars at greater orbital distances \citep{Kipping_2016}. Nonetheless, recent advances in observational techniques have enabled the detection and characterization of increasingly smaller (and potentially rocky) planets \citep{espresso,plato}.

Most detected rocky planets are Super-Earths, with radii below $2~R_\oplus$ and masses between 1 and $10~M_\oplus$, while planets smaller than Earth remain rare in current catalogs due to detection challenges \citep{Cloutier_2024}. Rocky planets are separated from gas-rich sub-Neptunes by the radius valley (`Fulton gap') at 1.5--2.0~$R_\oplus$ \citep{Fulton_2017,Fulton_2018,Petigura_2022}. The valley is attributed to atmospheric escape through photoevaporation \citep{Owen_2017,Venturini_2020,Burn_2024} or core-powered mass loss \citep{Gupta_2019,Gupta_2020,Gupta_2021,Rogers_2021}, while sub-Neptunes retain their primary atmospheres.

One of the main open questions regarding rocky exoplanets concerns their bulk elemental compositions and how they compare to those of their host stars. Addressing this helps constrain their mineralogy and internal structure. Planetary interiors play a key role in shaping atmospheric composition through outgassing processes \citep{Herbort_2020,Ortenzi_2020,Wordsworth_2022}. Understanding the interiors of the planet is therefore crucial for assessing its potential habitability \citep{Shahar_2019}.

Since we cannot directly measure the composition of rocky exoplanets, we must infer it by comparing observed bulk densities to interior structure models. These models typically assume a two-layer structure consisting of a metallic core and a silicate mantle, analogous to Earth, with iron (Fe) concentrated in the core and silicon (Si) residing mainly in the mantle. A first-order descriptor of a rocky planet's bulk composition is therefore the core mass fraction (CMF), the fraction of the planet's mass contained in the metallic core. Solar System terrestrial planets show wide CMF variation: Mars has the lowest ($\sim$0.232--0.241), followed by Earth and Venus ($\sim$0.325 and $\sim$0.31), with Mercury the highest (0.732--0.738) \citep{Rodr_guez_Mozos_2022}. Earth's core contains mainly Fe and nickel (Ni), while the mantle comprises oxygen (O; 44\%), Si (21\%), magnesium (Mg; 22\%), and minor Fe (6\%) \citep{Wang2018}.

In the Solar System, CI chondrites match the solar photospheric composition (except for highly volatile elements such as H, He, and noble gases) and are considered primitive remnants of the protoplanetary disk. 
In contrast, Earth's elemental abundances with respect to the CI chondritic (solar-like) composition follow a devolatilization trend: elements with a condensation temperature ($T_C$) less than $\sim 1200$ K (i.e. moderately volatile elements) are depleted relative to solar abundances, while those with a $T_C\gtrsim 1400$ K (i.e. refractory elements) are statistically equivalent between the two \citep{Wang2018,Wang2019,wang2019b, Mcdonough_2025}. The group of main elements (Fe, Mg, and Si) with an intermediate $T_C$ of $\sim$ 1200--1400 K are slightly depleted at the level of $\sim 10-20\%$ (relative to the major refractory element, Al; \citealt{Wang2019, wang2019b}) in the case of Earth.

For exoplanets, the bulk planetary Fe/Si ratio is often assumed to be similar to that of the host star \citep{Thiabaud_2015, santos15_01_of_02_unique_01_01_2025,Brugger_2017, dorn2015, Plotnykov_2020, acuna2021}. 
However, it remains debatable whether observational data provide evidence for this assumption. Several studies of rocky exoplanets found that their CMF-derived Fe/Si ratios tend to be higher than the photospheric Fe/Si of their host stars \citep{Adibekyan_2021, schulze_2021, liu_and_ni2023, chen26}, while \cite{brinkman2024a, brinkman2024b, plotnykov26} find a 1:1 relationship in the planet-star Fe/Si ratios. Explanations for the non-1:1 trend involve mantle stripping or late-stage collisions \citep{Scora_2020, scora2022, ferich25}, formation of super-Mercuries in the vicinity of rocklines \citep{Aguichine_2020}, and/or Fe nucleation \citep{johansen_dorn2022}. However, higher Fe/Si ratios do not necessarily indicate Fe enrichment; they could instead reflect Si depletion caused by devolatilization processes \citep{Wang2019, wang2019b}. We note that devolatilization and collisional mechanisms are not mutually exclusive; in this work we assess only whether devolatilization can account for moderate, Earth-like departures from a 1:1 planet--star refractory ratio, not for the extreme iron enrichment of Mercury-like planets.

In this work, we derive the bulk composition of rocky planets and compare them to their host stars' elemental ratios, exploring whether devolatilization can explain the observed stellar-planetary Fe/Si relationship. In Section~\ref{sec:MSEI} we describe the two forward models and the precomputed grid used to run the retrievals. In Section~\ref{sec:InteriorCompositionRetrieval} we outline the Bayesian framework and MCMC implementation. We present the sample in Section~\ref{sec:sample}. We show the results, including the sensitivity analysis and devolatilization trends, in Section~\ref{sec:results}. Finally, in Section~\ref{sec:discussion} we discuss the implications of our findings and prospects for future work.

\label{sec:intro}

\section{Planet interior models}

\subsection{The Marseille Super-Earth Interior model}
\label{sec:MSEI}

\begin{figure}[H]
    \centering
    \includegraphics[width=1\linewidth]{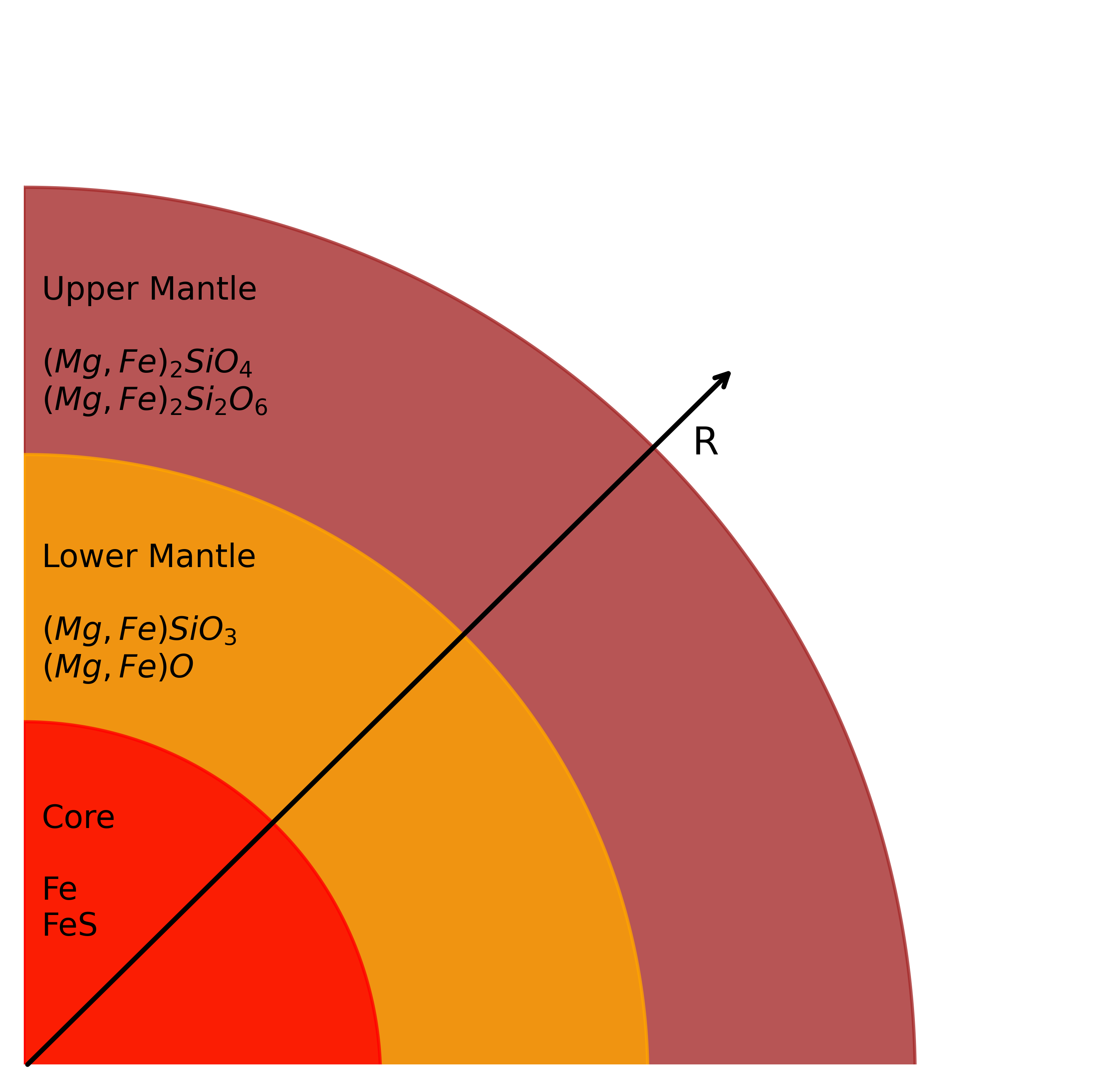}
    \caption{Schematic view of the interior model.}
    \label{fig:interior_model}
\end{figure}

The Marseille Super-Earth Interior Model \citep[MSEI;][]{brugger2016, Brugger_2017, brugger2018} allows us to simulate the interior of rocky exoplanets up to \mbox{$\sim10$} Earth masses. It assumes a fully differentiated, spherically symmetric concentric layer structure. The innermost layer is a liquid core made of Fe and FeS. Around it, there is a lower mantle of Bridgmanite \((\text{Mg},\text{Fe})\text{SiO}_3\) and Periclase \((\text{Mg},\text{Fe})\text{O}\), and surrounding that, an upper mantle of olivine \((\text{Mg},\text{Fe})_2 \text{SiO}_4\) and ortho-pyroxene \((\text{Mg},\text{Fe})_2 \text{Si}_2 \text{O}_6\). MSEI also accommodates liquid water and ice layers. Given the high bulk densities of the planets in our sample, together with their short orbital distances and high equilibrium temperatures, we assume that volatiles with low condensation temperatures, such as H$_{2}$O, are absent. Small planets on short-period orbits are expected to have lost any primordial gas envelope through photoevaporation \citep{Owen_2017, Lopez_2017}. Therefore, in this work, we set the water mass fraction to zero in all our models and only consider the existence of the core, lower mantle and upper mantle layers. This is nonetheless an assumption. If a planet retained a thin volatile envelope, a denser interior would be needed to reproduce its mass and radius, so our CMF would be a lower limit. A diagram of the interior structure model is displayed in Fig. \ref{fig:interior_model}.

Four parameters determine the composition of the planet: CMF, which represents the fraction of the mass contributed by the core to the total mass of the planet; \(f_{alloy}\), which is the fraction of iron-sulfur (S) alloy in the core; the number of moles of Mg relative to the number of moles of Si of the planet \((Mg/Si)_P\); and \(Mg\# = \frac{Mg}{Mg+Fe}_{mantle}\), which represents the amount of iron present in the silicate mantle and defines the level of differentiation of the planet \citep{sotin2007}. MSEI considers the mantle to be chemically homogeneous, meaning the Mg/Si and Fe/Si ratios are the same in the upper and lower mantle. 
The rest of the parameters are the mass of the planet ($M_{\rm p}$) and the temperature ($T_{\rm surf}$) and pressure ($P_{\rm surf}$) at the surface level. For this work, the $T_{\rm surf}$ was fixed to 1500 K and $P_{\rm surf}$ to $10^5$ Pa. The $T_{\rm surf}$ and $P_{\rm surf}$ have been shown to produce changes of less than 1\% in planet radius at a constant composition and mass \citep{brugger2016}.

For a given set of input parameters, MSEI integrates the profiles of gravitational acceleration, density, pressure, and temperature over a discrete one-dimensional radial grid, returning the planetary radius ($R_p$) and the elemental mole fractions of iron, magnesium, silicon, oxygen, and sulfur ($N_{\rm Fe}$, $N_{\rm Mg}$, $N_{\rm Si}$, $N_{\rm O}$, $N_{\rm S}$).

The radial profile of gravitational acceleration, $g(r)$, follows from Gauss's law (Eq.~\ref{eq:gr_mass}),
\begin{equation}
\frac{dg}{dr} = 4\pi G\rho - \frac{2Gm}{r^3}, \label{eq:gr_mass}
\end{equation}
where $G = 6.674 \times 10^{-11}~\mathrm{m^3\,kg^{-1}\,s^{-2}}$ is Newton's gravitational constant and $m(r)$ is the enclosed mass at radius $r$, given by mass conservation (Eq.~\ref{eq:mass_integral}),
\begin{equation}
\frac{dm}{dr} = 4\pi r^2 \rho, \label{eq:mass_integral}
\end{equation}
with $\rho(r)$ the density. The pressure profile is determined by hydrostatic equilibrium (Eq.~\ref{eq:hydrostatic_equilibrium}),
\begin{equation}
\frac{dP}{dr} = -\rho g. \label{eq:hydrostatic_equilibrium}
\end{equation}
The temperature gradient within each layer is computed under the assumption of adiabatic convection,
\begin{equation}
\frac{dT}{dr} = -g \frac{\gamma T}{\Phi}, \quad
\begin{cases}
\gamma = \gamma_0 \left(\frac{\rho_0}{\rho}\right)^q, \\
\Phi = \frac{\mathscr{K}}{\rho} = \frac{dP}{d\rho},
\end{cases}
\label{eq:temperature_gradient}
\end{equation}
where $\gamma_0$ is the reference Grüneisen parameter, $\rho_0$ the reference density, $q$ the adiabatic power exponent \citep{brugger2018}, and $\mathscr{K}$ the bulk modulus. Thin conductive layers at the core-mantle boundary are not explicitly resolved; instead, following \cite{sotin2007}, their thickness is set to zero and a fixed temperature jump of 800~K is imposed at the corresponding radial position, consistent with the terrestrial core-mantle boundary value \citep{anderson1982, spiliopoulos1984}.

Density is computed numerically from the Vinet equation of state \citep{vinet1987} (Eq.~\ref{eq:EOS}), which has been widely adopted for modeling exoplanetary interiors \citep{unterborn2016, seager2007, valencia2007c},
\begin{equation}
P = 3K_0 \left[\left(\frac{\rho}{\rho_0}\right)^{\frac{2}{3}} - \left(\frac{\rho}{\rho_0}\right)^{\frac{1}{3}}\right] \exp\left\{\frac{3}{2}(K'_0-1)\left[1 - \left(\frac{\rho}{\rho_0}\right)^{-\frac{1}{3}}\right]\right\}. \label{eq:EOS}
\end{equation}
The reference parameters $\rho_0$, $K_0$, and $K'_0$ are listed in Tables 2.2 and 2.3 of \cite{brugger2018}. The model iterates over $g(r)$, $P(r)$, and $T(r)$ using updated density values from the EoS until convergence is reached, defined as a relative change below $10^{-5}$ between successive iterations with no displacement of layer boundaries on the grid.

\subsection{The pyExoInt model}
\label{sec:pyExoInt}
For a detailed description of \texttt{pyExoInt}, we refer to \cite{Wang2019, Wang2022}. Briefly, the model adopts a chemical system of major oxides and reduced phases (Na$_2$O--CaO--MgO--Al$_2$O$_3$--SiO$_2$--FeO--NiO--SO$_3$--CO$_2$--C--metals) for the mantle, and an Fe--Ni--Si--S alloy for the core of a terrestrial-type exoplanet. CMF is determined by mass balance through partitioning the bulk planetary composition into stoichiometric mantle and core components.

Given the resulting compositions and CMF, planetary mineralogy and internal structure are computed using the Mineral module of \texttt{pyExoInt}, which incorporates the \texttt{PERPLE\_X} framework \citep{Connolly2009}. Mantle mineralogy is obtained via Gibbs free energy minimization based on bulk composition and pressure--temperature conditions. Thermodynamic parameters and equations of state are adopted from \cite{Stixrude2011} and \cite{Kuwayama2020}.

An adiabatic thermal profile is assumed, with a mantle potential temperature of 1700~K at 1~bar, broadly representative of Earth-like conditions. Variations in mantle potential temperature primarily affect mineralogy in the transition zone, while having a limited influence on the overall internal structure \citep{Hinkel2018, Wang2022}. A temperature jump is also imposed at the core--mantle boundary following \cite{Noack2020}.

\subsection{Grid of interior structure models}
\label{sec:grid}

As described in Sections \ref{sec:MSEI} and \ref{sec:pyExoInt}, both forward models take the compositional parameters $\theta = (CMF, f_{alloy}, Mg\#, \frac{Mg}{Si})$ as input, but differ in their treatment of the planetary observables: MSEI takes $M_p$ as input and returns $R_p$, while pyExoInt takes $R_p$ as input and returns $M_p$. To describe the retrieval framework in a unified way, we define $X_p$ as the observable input to the forward model and $Y_p$ as the predicted observable output, such that $Y_p = \mathcal{F}(X_p, \theta)$. Both models additionally return the elemental mole fractions $N_{Fe}$, $N_{Mg}$, $N_{Si}$, $N_{O}$, and $N_{S}$.

Computing the radius and elemental mole fractions for a single set of input parameters with MSEI or pyExoInt takes in the order of 10 seconds, making it prohibitively expensive to evaluate on-the-fly within a retrieval that requires thousands of forward model calls. To address this, we precomputed a six-dimensional grid spanning the range of rocky exoplanet compositions, from which the radius and elemental mole fractions are obtained by interpolation at any input parameters within the grid domain.

\begin{figure}
        \centering
        \includegraphics[width=1\linewidth]{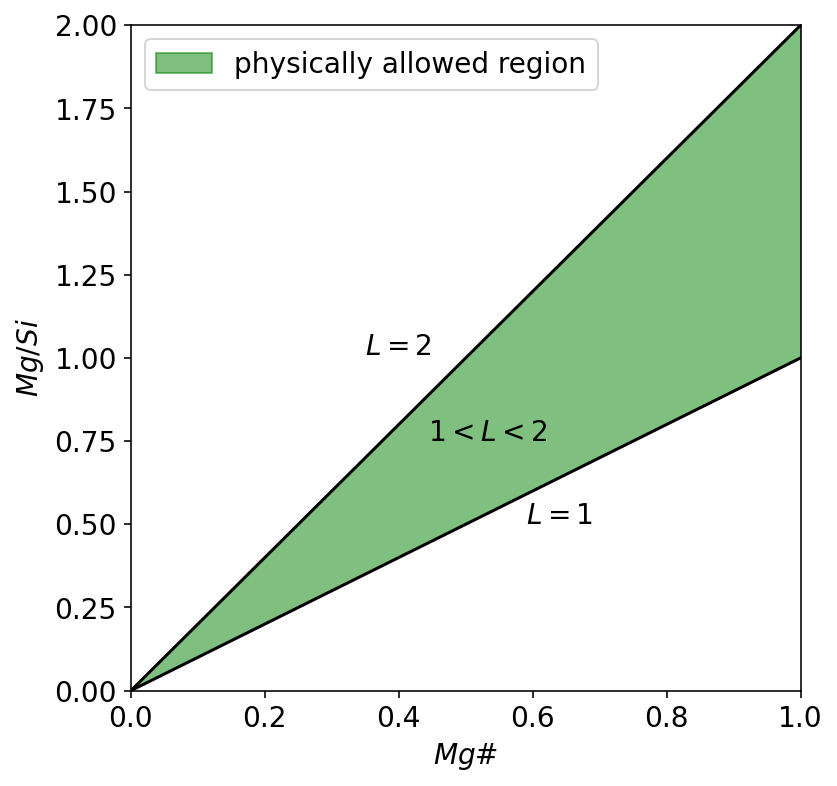}
        \caption{Physically allowed region in the $Mg\#,Mg/Si$ plane, as defined by the upper and lower limit values of the slope parameter L.}
        \label{fig:L_plot}
    \end{figure}

In MSEI, the constraint \(2 \, \text{Mg\#} > \text{Mg/Si} > \text{Mg\#}\) \citep{Brugger_2017} must be satisfied by construction of the model. Imposing this condition on the grid parameters, however, would produce a non-rectangular grid, which is impractical for interpolation. We avoid this by introducing the reparameterization \(L = \frac{\text{Mg/Si}}{\text{Mg\#}}\), under which the constraint reduces to $1 < L < 2$. This also simplifies the MCMC priors (see Sect.~\ref{sec:InteriorCompositionRetrieval}). The allowed region in the $Mg\#$ -- $\mathrm{Mg/Si}$ plane under the new parameterization is illustrated in Fig.~\ref{fig:L_plot}, and the grid node values are listed in Table~\ref{tab:grid-range}. 

To quantify the precision of the interpolation, we drew $N = 100$ planets with randomly sampled $X_{\rm p,k}$ within the limits of the model, where $k$ indexes the planet, computed $Y_{\rm p,k}$ directly with MSEI and pyExoInt, and compared the results to the respective interpolated values (Eq.~\ref{eq:interp_err}). The resulting interpolation errors are below 1\%, well within the observational radius and mass uncertainties of our sample (see Sect.~\ref{sec:sample}), confirming that the grid is sampled finely enough to yield a smooth and accurate representation of the forward models.

\begin{equation}
    \text{err}^\%_{\text{interpolation}} = \frac{100}{N} \sum_{k=1}^{N} \left|\frac{Y_p(X_{p,k})^{\text{interpolated}} - Y_p(X_{p,k})^{\text{model}}}{Y_p(X_{p,k})^{\text{model}}}\right|
    \label{eq:interp_err}
\end{equation}

    \begin{center}
    \begin{table}[h!] 
    \centering
    \caption{Parameter values used to calculate the grid of forward models.} 
    \label{tab:grid-range} 
    \begin{tabular}{lc}
    \hline
    \hline
    Parameter & Range \\ \hline
    \(M_{\text{P}}\) & \(0.5 \leq M_{\text{P}} \leq 10.2, \, \Delta M_{\text{P}} = 0.2\) \\ 
    \(CMF\) & \(0.0 \leq CMF \leq 1.0, \, \Delta CMF = 0.05\) \\ 
    \(\text{Mg\#}\) & \(\{0.1, 0.4, 0.7, 1.0\}\) \\ 
    \(\text{L}\) & \(\{1.01, 1.25, 1.5, 1.75, 2.0\}\) \\ 
    \(f_{\text{alloy}}\) & \(\{0.0, 0.25, 0.5, 0.75, 1.0\}\) \\ \hline
    \end{tabular}
    \end{table}
    \end{center}

We calculate and store each point of the grid in an HDF5 file. To interpolate, we use the \verb|RegularGridInterpolator| class from the \verb|Scipy| python library which performs multidimensional linear interpolation over a hyper-rectangular grid. As an example, we show the variation of $R_p$, in the $M_p$-$CMF$ plane for the MSEI interpolated forward model in Fig.~\ref{fig:R(mass,cmf)}.

\begin{figure}
    \centering
    \includegraphics[width=1\linewidth]{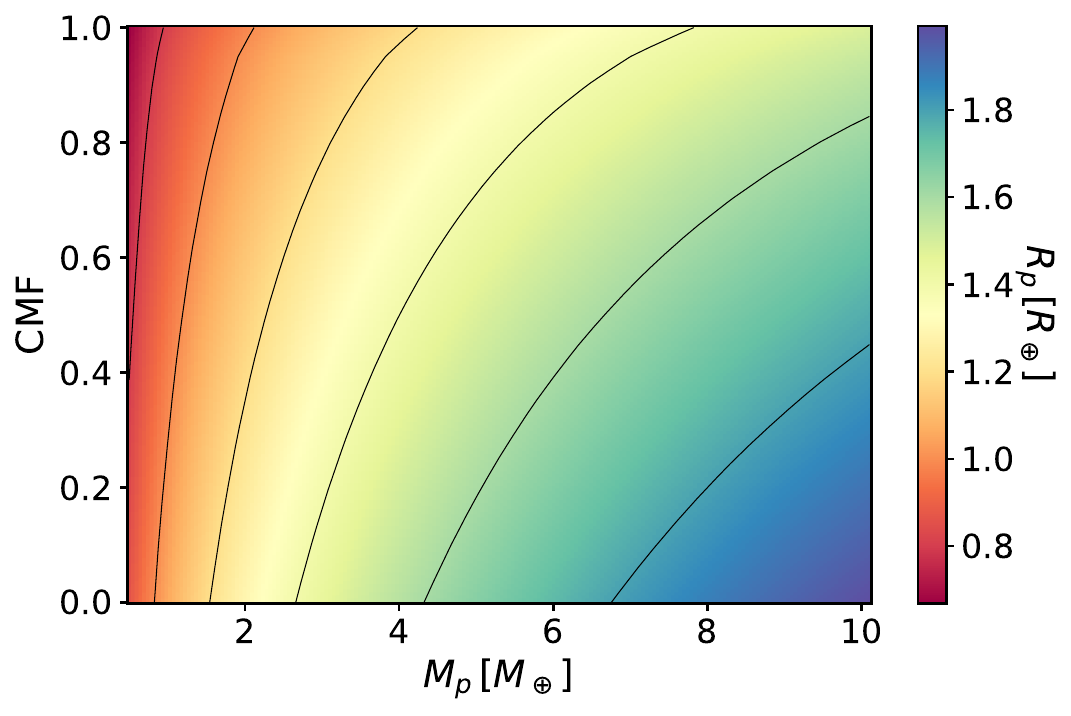}
    \caption{2-dimensional slice of the interpolated forward model grid showing that the radius is a smooth function of the mass and the \( CMF \). }
    \label{fig:R(mass,cmf)}
\end{figure}

\section{Interior composition retrievals}
\label{sec:InteriorCompositionRetrieval}

To determine the internal composition of the planet, we estimate the non-observable parameters $\theta$ using only the observed quantities $X_{obs}$ and $Y_{obs}$, together with their uncertainties. This requires solving the inverse problem: estimating the probability distribution of $\theta$ given the observations \citep{Mosegaard1995}. This is accomplished via Bayes' theorem, which relates the posterior $P(\theta \mid X_{obs}, Y_{obs})$ to the prior $P(\theta)$, the likelihood $P(X_{obs}, Y_{obs} \mid \theta)$, and the marginal likelihood $P(X_{obs}, Y_{obs})$ (Eq.~\ref{eq:bayes}):

\begin{equation}
    P(\theta \mid X_{obs}, Y_{obs}) = \frac{P(X_{obs}, Y_{obs} \mid \theta)\, P(\theta)}{P(X_{obs}, Y_{obs})}.
    \label{eq:bayes}
\end{equation}

The likelihood quantifies how well a given parameter set explains the observations. In practice, we work with the log-likelihood $\mathcal{L}(\theta \mid X_{obs}, Y_{obs})$ (Eq.~\ref{eq:log_likelihood}), as the logarithm is monotonic and avoids floating-point overflow from large probability values:
\begin{equation}
\mathcal{L}(\theta \mid X_{obs}, Y_{obs}) = \ln P(X_{obs}, Y_{obs} \mid \theta).
\label{eq:log_likelihood}
\end{equation}
The marginal likelihood $P(X_{obs}, Y_{obs})$ is a normalization constant that does not affect the moments of the posterior and is set to unity.

The explicit form of the log-likelihood is given by Eq.~\ref{eq:log_likelihood_full} \citep{acuna2021, dorn2015}:

\begin{equation}
\begin{split}
\mathcal{L}(\theta \mid X_{\text{obs}}, Y_{\text{obs}},
\sigma_{X+}, \sigma_{X-}, \sigma_{Y+}, \sigma_{Y-}) = \\
-\frac{1}{2} \left[ \left( \frac{X_p - X_{\text{obs}}}{\sigma_{X}} \right)^2 + \left( \frac{Y_p(X_p,\theta) - Y_{\text{obs}}}{\sigma_{Y}} \right)^2 \right],
\end{split}
\label{eq:log_likelihood_full}
\end{equation}

where:
\begin{align*}
&Y_p(X_p,\theta), \quad \text{is the observable predicted by the forward model},\\
&X_p, \quad \text{is the input observable sampled from the prior distribution},\\
&\theta, \quad \text{denotes the parameters } CMF,\, f_{alloy},\, Mg\#,\, \tfrac{Mg}{Si} \text{ sampled}\\
&\quad\quad\text{from the prior distribution},\\
&X_{\text{obs}},\, Y_{\text{obs}}, \quad \text{are the observed planetary quantities},\\
&\sigma_{X+},\, \sigma_{X-},\, \sigma_{Y+},\, \sigma_{Y-}, \quad \text{are the observational uncertainties},\\
&\sigma_{X} =
\begin{cases}
\sigma_{X+}, & \text{if } X_p > X_{\text{obs}}, \\
\sigma_{X-}, & \text{if } X_p \leq X_{\text{obs}},
\end{cases}\\
&\sigma_{Y} =
\begin{cases}
\sigma_{Y+}, & \text{if } Y_p > Y_{\text{obs}}, \\
\sigma_{Y-}, & \text{if } Y_p \leq Y_{\text{obs}}.
\end{cases}
\end{align*}

For MSEI, $X_p = M_p$ and $Y_p = R_p$; for pyExoInt, $X_p = R_p$ and $Y_p = M_p$.

For the prior distributions of $X_p$ and $\theta$, we adopt uniform distributions spanning the full grid domain, as detailed in Section~\ref{sec:HD15337b}, Table~\ref{tab:prior_distributions}. Although a Gaussian prior on $X_p$ centered on $X_{obs}$ could in principle be used, a uniform prior is equivalent here because the observational uncertainty is already incorporated in the log-likelihood. Under these conditions, the effective posterior on $X_p$ converges to a Gaussian \citep{gelman2013bayesian}. The uniform prior is simpler to implement, and any gain in convergence speed from a Gaussian prior is negligible given the computational efficiency of the interpolation-based forward model.

Since the posterior cannot be evaluated analytically, we sample it using the affine-invariant MCMC ensemble sampler of \cite{10.2140/camcos.2010.5.65}, implemented in the \texttt{emcee} package \citep{2013PASP..125..306F}. Convergence is assessed by verifying that the autocorrelation time $\tau$ satisfies $\tau < n/50$, where $n$ is the number of steps in the chain. Another parameter required by \texttt{emcee} is the number of chains. In this work, we used 32.

\section{Sample selection}
\label{sec:sample}

\begin{figure}
    \centering
    \includegraphics[width=1\linewidth]{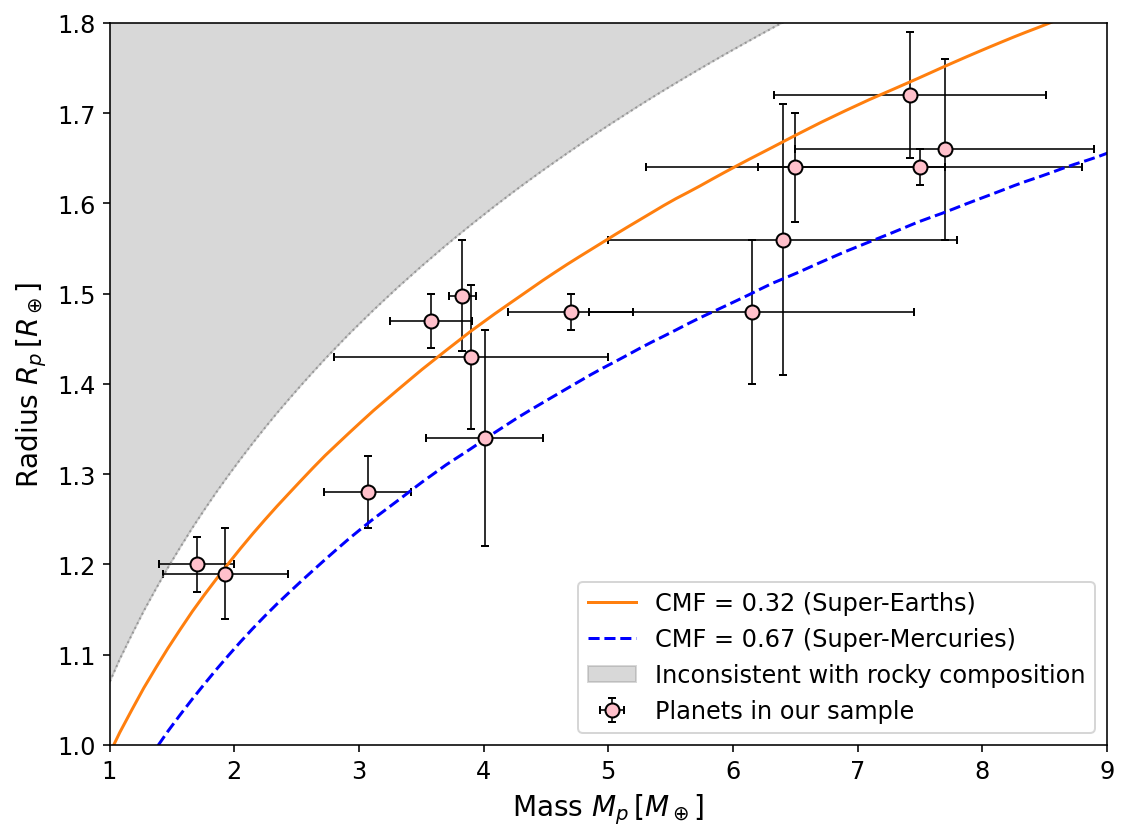}
\caption{Mass-radius diagram of the planets in our sample with respective errors. For comparison purposes we show MSEI forward models for CMF = 0.32 and CMF = 0.67, representing Super-Earths and Super-Mercuries respectively. The gray shaded region indicates the part of the mass-radius diagram corresponding to compositions less dense than the pure-mantle threshold, the lowest-density composition possible for a rocky planet. By selecting planets outside this gray area, we rule out planets whose density is inconsistent with a rocky interior without a gas layer (e.g. H$_{2}$O, H$_{2}$).}
    \label{fig:MR_diagram_Sample}
\end{figure}

We searched in the NASA Exoplanet Archive\footnote{\url{https://exoplanetarchive.ipac.caltech.edu/}} for rocky planets with $0.5 \ M_{\oplus} < M_p < 10 \ M_{\oplus}$ and $0.5 \ R_{\oplus} < R_p < 2 \ R_{\oplus}$ with errors of the mass and radius below 30\% of the corresponding value, a level at which the bulk composition of a planet can still be meaningfully constrained \citep[see Fig.~6 of][]{Otegi_2020}. We choose the mass-radius curve in our forward model with the composition that corresponds to the lowest density. This composition has no iron core (CMF = 0) and no iron in the mantle (Mg\# = 1), making it the least dense interior a volatile-free rocky planet can have. Planets below this curve are too dense to require a significant volatile layer and are therefore likely rocky. We then select only planets that are below this curve in the mass-radius diagram (see Fig.~\ref{fig:MR_diagram_Sample}). This criterion also takes into account the 1$\sigma$ uncertainty in radius, $R_{model} =R_{obs}+\sigma_{R_{obs}}$, where $R_{model}$ is the mass-radius curve of the lowest density for a rocky planet: CMF = 0, $f_{alloy}=1$, Mg\# = 1 and Mg/Si = 1.131. With this criterion, we minimize the possibility that the final sample contains planets with substantial secondary envelopes, which could bias our inference of the core mass fraction and elemental ratios \citep{Plotnykov_2020}.

Equilibrium temperatures were computed homogeneously for the full sample from the orbital period and the stellar mass, radius, and effective temperature, assuming zero albedo (see Appendix~\ref{sec:equilibrium_temp_appendix}). Where a previously published value exists, it is listed alongside our computed one in Table~\ref{tab:planet_parameters} for comparison. The equilibrium temperature is used only in the correlation analysis of Fig.~\ref{fig:corr} and does not enter the interior retrieval.

Stellar abundances of Fe, Mg, Si, O, and S were obtained from the Hypatia catalog \citep{hypatia}, which at the time of its release\footnote{December 2017. https://www.hypatiacatalog.com/} compiled spectroscopic data from 84 literature sources for 50 elements across 3058 stars in the solar neighborhood. Since Hypatia compiles many independent catalogs, for each element and planet we use the latest published abundance and normalize it to the solar composition of \cite{Asplund_2009}. For abundances without reported uncertainties, we adopt the median error across the contributing catalogs as listed in Hypatia. We retained only planets whose host star had measured abundances for at least Fe, Mg, Si, and O. Our final sample consists of 14 planets, of which only 5 have the full set of abundances including S; 8 host stars lack S measurements, and 1 (TOI-431) lacks both S and O. All stellar abundances and uncertainties are listed in Table~\ref{tab:stellar_parameters}, and planetary parameters are given in Table~\ref{tab:planet_parameters}. A mass-radius diagram of the full sample is shown in Fig.~\ref{fig:MR_diagram_Sample}.

\begin{table*}[]
    \centering
    \caption{Mass, radius, and equilibrium temperature of the exoplanets in our sample.}
    \label{tab:planet_parameters}
    \begin{tabular}{llllr}
    \hline
    \hline
        Planet ID &$ R_p [R_{\oplus}]$ & $M_p [M_{\oplus}]$ & $T_{eq} [K]$ & Reference \\ \hline \\
        Kepler-407 b & $1.19 \pm 0.05$ & $1.93 \pm 0.5$ & $2214 \pm 69$ & \cite{brinkman2024b} \\ \\
        Kepler-406 b & $1.56 \pm 0.15$ & $6.4 \pm 1.4$ & $1425 \pm 39$ & \cite{brinkman2024a} \\ \\
        Kepler-78 b & $1.2 \pm 0.03$ & $1.7 \pm 0.3$ & $2207 \pm 39$ & \cite{brinkman2024a} \\
        &&&$2223 \pm 32$&\cite{Bonomo2023}\\ \\
        HD 20329 b & $1.72 \pm 0.07$ & $7.42 \pm 1.09$ & $2141 \pm 33$ & \cite{Murgas2022} \\
        &&&$2141 \pm 27$&\cite{Murgas2022}\\ \\
        Kepler-21 b & $1.64 \pm 0.02$ & $7.5 \pm 1.3$ & $1939 \pm 52$ & \cite{brinkman2024a} \\
        &&&$2025 \pm 20$&\cite{LopezMorales}\\ \\
        K2-38 b & $1.66 \pm 0.1$ & $7.7 \pm 1.2$ & $1289 \pm 34$ & \cite{brinkman2024a} \\
        &&&$1266^{+44}_{-50}$&\cite{toledo2020}\\ \\
        Kepler-99 b & $1.48 \pm 0.08$ & $6.15 \pm 1.3$ & $894 \pm 36$ & \cite{Marcy2014} \\ \\
        Kepler-36 b & $1.498^{+0.061}_{-0.049}$ & $3.83^{+0.11}_{-0.1}$ & $1074 \pm 15$ & \cite{Vissapragada2020} \\
        &&&$978 \pm 11$&\cite{carter2012}\\ \\
        K2-36 b & $1.43 \pm 0.08$ & $3.9 \pm 1.1$ & $1336 \pm 30$ & \cite{brinkman2024a} \\
        &&&$1328 \pm 12$&\cite{Bonomo2023}\\ \\
        Kepler-100 b & $1.34 \pm 0.12$ & $4.01 \pm 0.47$ & $1278 \pm 32$ & \cite{brinkman2024b} \\ \\
        Kepler-93 b & $1.48 \pm 0.02$ & $4.7 \pm 0.5$ & $1134 \pm 33$ & \cite{brinkman2024a} \\
        &&&$1133 \pm 17$&\cite{Bonomo2023}\\ \\
        TOI-431 b & $1.28 \pm 0.04$ & $3.07 \pm 0.35$ & $1889 \pm 49$ & \cite{osborn2021} \\
        &&&$1862 \pm 42$&\cite{osborn2021}\\ \\
        Kepler-10 b & $1.47 \pm 0.03$ & $3.58 \pm 0.33$ & $2112 \pm 75$ & \cite{brinkman2024b} \\ \\
        HD 15337 b & $1.64 \pm 0.06$  & $6.5 \pm 1.2$ & $984 \pm 21$ & \cite{brinkman2024a} \\
        &&&$1001 \pm 12$&\cite{dumusque2019}\\ \\\hline
    \end{tabular}
    \tablefoot{Masses and radii are taken from the first listed reference. The equilibrium temperature in the first row of each entry was computed from the orbital period and the stellar mass, radius, and effective temperature adopted from the Hypatia catalog \citep{hypatia}, assuming zero albedo (Appendix~\ref{sec:equilibrium_temp_appendix}), instead of being taken directly from the literature. Where available, the previously published temperature is listed in the second row, next to its reference. Differences between the two partly reflect differing albedo assumptions in the literature.}
\end{table*}

\begin{table*}[]
    \centering
    \caption{Stellar abundances for the host stars of the planets in our sample.}
    \label{tab:stellar_parameters}
\begin{tabular}{llllll} 
\hline\hline
Host       & [Fe/H] [dex]         & [Mg/H] [dex]         & [Si/H] [dex]         & [O/H] [dex]          & [S/H][dex]           \\ 
\hline
Kepler-407 & $0.34 \pm 0.04^{BR18}$  & $0.32 \pm 0.07^{BR18}$  & $0.32 \pm 0.02^{BR18}$  & $0.17 \pm 0.02^{BR18}$  & -                     \\
Kepler-406 & $0.22 \pm 0.02^{BR18}$  & $0.21 \pm 0.02^{BR18}$  & $0.2 \pm 0.02^{BR18}$   & $0.18 \pm 0.06^{BR18}$  & -                     \\
Kepler-78  & $-0.01 \pm 0.04^{BR16}$ & $-0.07 \pm 0.07^{BR16}$ & $-0.04 \pm 0.05^{BR16}$ & $-0.07 \pm 0.09^{BR16}$ & -                     \\
HD 20329   & $-0.01 \pm 0.05^{RA}$   & $0.037 \pm 0.07^{BE}$   & $0.022 \pm 0.05^{BE}$   & $0.06 \pm 0.04^{RA}$    & $0.014 \pm 0.09^{BE}$   \\
Kepler-21  & $-0.04 \pm 0.16^{SC}$   & $0.0 \pm 0.07^{SC}$     & $-0.07 \pm 0.12^{SC}$   & $-0.09 \pm 0.13^{SC}$   & $-0.17 \pm 0.03^{SC}$   \\
K2-38      & $0.22 \pm 0.04^{BR16}$  & $0.21 \pm 0.07^{BR16}$  & $0.21 \pm 0.05^{BR16}$  & $0.15 \pm 0.09^{BR16}$  & -                     \\
Kepler-99  & $0.26 \pm 0.04^{BR18}$  & $0.27 \pm 0.02^{BR18}$  & $0.16 \pm 0.04^{BR18}$  & $-0.03 \pm 0.08^{BR18}$ & -                     \\
Kepler-36  & $-0.18 \pm 0.04^{BR18}$ & $-0.18 \pm 0.07^{BR18}$ & $-0.19 \pm 0.01^{BR18}$ & $0.01 \pm 0.04^{BR18}$  & -                     \\
K2-36      & $-0.01 \pm 0.04^{BR16}$ & $-0.04 \pm 0.07^{BR16}$ & $0.03 \pm 0.05^{BR16}$  & $-0.06 \pm 0.09^{BR16}$ & -                     \\
Kepler-100 & $0.06 \pm 0.02^{LI}$    & $0.081 \pm 0.02^{LI}$   & $0.106 \pm 0.03^{LI}$   & $0.008 \pm 0.05^{LI}$   & $0.167 \pm 0.09^{LI}$   \\
Kepler-93  & $-0.11 \pm 0.04^{SA}$   & $-0.1 \pm 0.04^{SA}$    & $-0.21 \pm 0.03^{SA}$   & $-0.167 \pm 0.04^{LI}$  & $-0.114 \pm 0.09^{LI}$  \\
TOI-431    & $0.08 \pm 0.02^{SU}$    & $0.06 \pm 0.07^{AD}$    & $0.073 \pm 0.05^{AD}$   & -                     & -                     \\
Kepler-10  & $-0.1 \pm 0.05^{SA}$    & $0.03 \pm 0.1^{SA}$     & $-0.11 \pm 0.03^{SA}$   & $0.16 \pm 0.1^{SA}$     & -                     \\
HD 15337   & $0.12 \pm 0.1^{SU}$     & $0.14 \pm 0.05^{SO}$    & $0.12 \pm 0.04^{SO}$    & $-0.01 \pm 0.08^{PO}$   & $0.46 \pm 0.09^{LU}$    \\
\hline
\end{tabular}
    \tablefoot{Abundances are the latest values found in the Hypatia catalog \citep{hypatia}. The reference of each abundance is denoted by the following keys:
LU = \cite{luck18_raw_01_01_2025}\tablefootmark{*},
PO = \cite{polanski22_raw_01_01_2025},
BR16 = \cite{brewer16_01_of_02_unique_01_01_2025}\tablefootmark{**},
SA = \cite{santos15_01_of_02_unique_01_01_2025},
BR18 = \cite{brewer18_01_of_02_unique_01_01_2025},
LI = \cite{liu20_01_of_02_unique_01_01_2025}\tablefootmark{**},
AD = \cite{adibekyan12_raw_01_01_2025},
BE = \cite{bedell17_raw_01_01_2025},
SO = \cite{soto18_02_of_04_unique_01_01_2025},
SC = \cite{schuler15_raw_01_01_2025},
SU = \cite{suarezandres17_raw_01_01_2025},
RA = \cite{ramirez13_raw_01_01_2025}.\\
\tablefoottext{*}{Catalogs that do not report uncertainties; values were therefore adopted from the standard errors in the Hypatia Catalog.}
\tablefoottext{**}{Catalogs with homogeneous uncertainties, i.e., the same uncertainty for all stars for a given element.}}
\end{table*}

\section{Results}
\label{sec:results}
\subsection{Forward model chemical sensitivity}
\label{sec:sensitivity}

\begin{figure*}
    \centering
    \includegraphics[width=\textwidth]{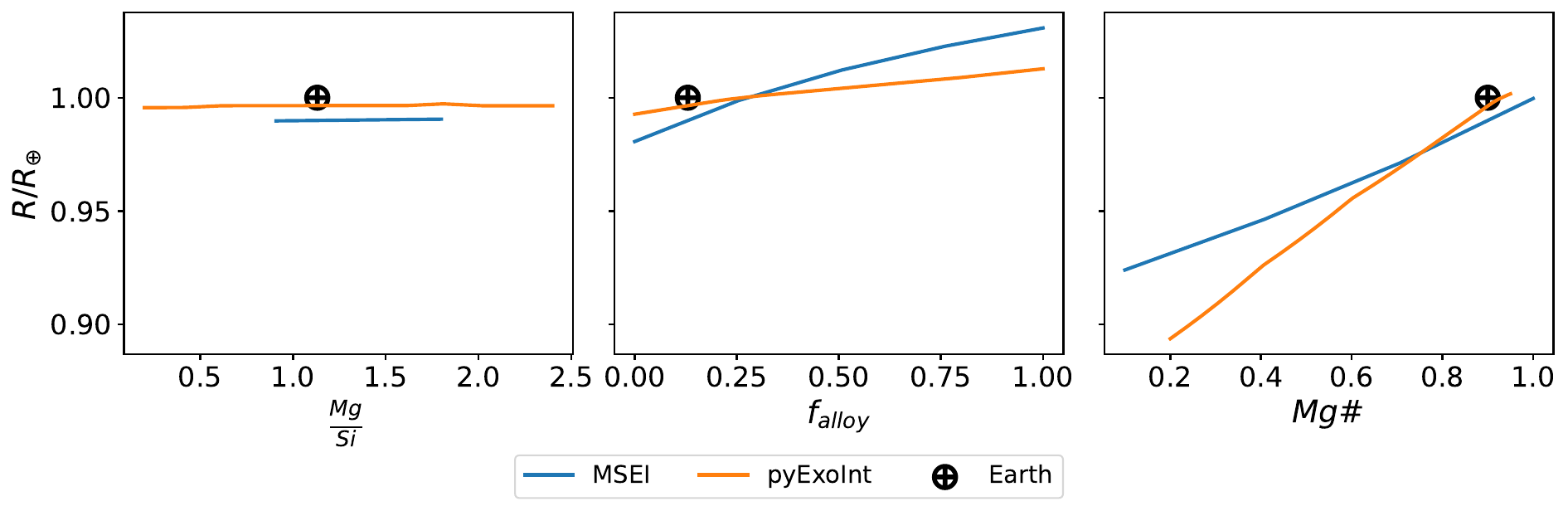}
    \caption{Predicted planetary radius $R/R_\oplus$ as a function of compositional parameters for an Earth-mass planet ($M = 1\, M_\oplus$) with Earth-like CMF. Each panel varies one parameter while holding the others fixed at Earth values (Mg/Si $= 1.131$, $f_\mathrm{alloy} = 0.13$, $Mg\# = 0.9$). From left to right: Mg/Si ratio, iron alloy fraction $f_\mathrm{alloy}$, and mantle magnesium number $Mg\#$. Results from two interior structure models are shown: MSEI (blue) and pyExoInt (orange). The Earth ($\oplus$) is marked in each panel.}
    \label{fig:chemical_curves}
\end{figure*}

To assess the sensitivity of the predicted radius to chemical composition, we evaluate both interior forward models for an Earth-mass planet ($M = 1\,M_\oplus$) with Earth-like CMF, varying $Mg/Si$, $Mg\#$ and $f_{alloy}$ across its allowed ranges while fixing the other two at Earth values. The results are shown in Fig.~\ref{fig:chemical_curves}.

The Mg/Si ratio has a negligible effect on the predicted radius in both models, making it a degenerate parameter for Earth-mass planets. The iron alloy fraction $f_\mathrm{alloy}$ has a moderate impact: MSEI predicts a radius variation from $\sim 0.98$ to $\sim 1.03\,R_\oplus$ (5\%) over the full $f_\mathrm{alloy}$ range, while pyExoInt predicts a narrower range of $\sim 0.99$ to $\sim 1.01\,R_\oplus$ (2\%). The magnesium number $Mg\#$ has the largest influence on the predicted radius, with MSEI yielding a variation from $\sim 0.92$ to $\sim 1.00\,R_\oplus$ (8\%), while pyExoInt spans a wider range of $\sim 0.89$ to $\sim 1.00\,R_\oplus$ (11\%).

\subsection{HD 15337 b as science case}
\label{sec:HD15337b}
Building on the Bayesian framework outlined in Section \ref{sec:InteriorCompositionRetrieval}, we define the prior distributions and present the first interior retrieval for HD 15337 b ($R = 1.64 \pm 0.06 \ R_{\oplus}$, $M = 6.5 \pm 1.2 \ M_{\oplus}$). We consider two retrieval configurations: a `simple' retrieval in which only the planet mass and CMF are free parameters with uniform priors, while $f_{alloy}$, $Mg\#$, and $\mathrm{Mg/Si}$ are fixed to Earth-like values ($f_{alloy} = 0.13$, $Mg\# = 0.9$, and $Mg/Si = 1.131$; \citealt{sotin2007, Stacey_2005}); and a `full' retrieval in which all five parameters are free with uniform priors spanning the full domain of the forward model (see Table \ref{tab:prior_distributions}). HD 15337 b serves as a benchmark for this first retrieval, as it combines relatively precise mass and radius measurements with mean values that lie well within the interior of our sample distribution.

\begin{table}[h]
\centering
\caption{Prior distribution parameters for the ``simple'' and ``full'' retrievals.}
\label{tab:prior_distributions}
\begin{tabular}{lcc}
\hline
\hline
Parameter & Prior$_{simple}$ & Prior$_{full}$\\ \hline
\multicolumn{3}{c}{MSEI} \\ \hline
\( M_{\text{P}} \) & \( \mathcal{U}(0.5, 10.2) \) & \( \mathcal{U}(0.5, 10.2) \) \\
\( CMF \) & \( \mathcal{U}(0.0, 1.0) \) & \( \mathcal{U}(0.0, 1.0) \) \\
\( \text{Mg\#} \) & 0.9 & \( \mathcal{U}(0.1, 1.0) \) \\
\( L \) & 1.257 & \( \mathcal{U}(1.01, 2.0) \) \\
\( f_{\text{alloy}} \) & 0.13 & \( \mathcal{U}(0.0, 1.0) \) \\ \hline
\multicolumn{3}{c}{pyExoInt} \\ \hline
\( R_{\text{P}} \) & \( \mathcal{U}(0.6, 1.6) \) & \( \mathcal{U}(0.6, 1.6) \) \\
\( CMF \) & \( \mathcal{U}(0.0, 0.9) \) & \( \mathcal{U}(0.0, 0.9) \) \\
\( \text{Mg\#} \) & 0.9 & \( \mathcal{U}(0.2, 0.8) \) \\
\( \text{Mg/Si} \) & 1.131 & \( \mathcal{U}(0.2, 2.4) \) \\
\( f_{\text{alloy}} \) & 0.13 & \( \mathcal{U}(0.0, 1.0) \) \\ \hline
\end{tabular}
\tablefoot{For both MSEI and pyExoInt, the simple model fixes $f_{alloy}$, $Mg\#$, and the silicon ratio to Earth values \citep{sotin2007, Stacey_2005}, while the full model treats them as free parameters with uniform priors. Note that MSEI uses the reparameterized quantity $L = Mg/Si\,/\,Mg\#$, whereas pyExoInt uses $Mg/Si$ directly.}
\end{table}

\begin{figure*}
    \centering
    \includegraphics[width=1\textwidth]{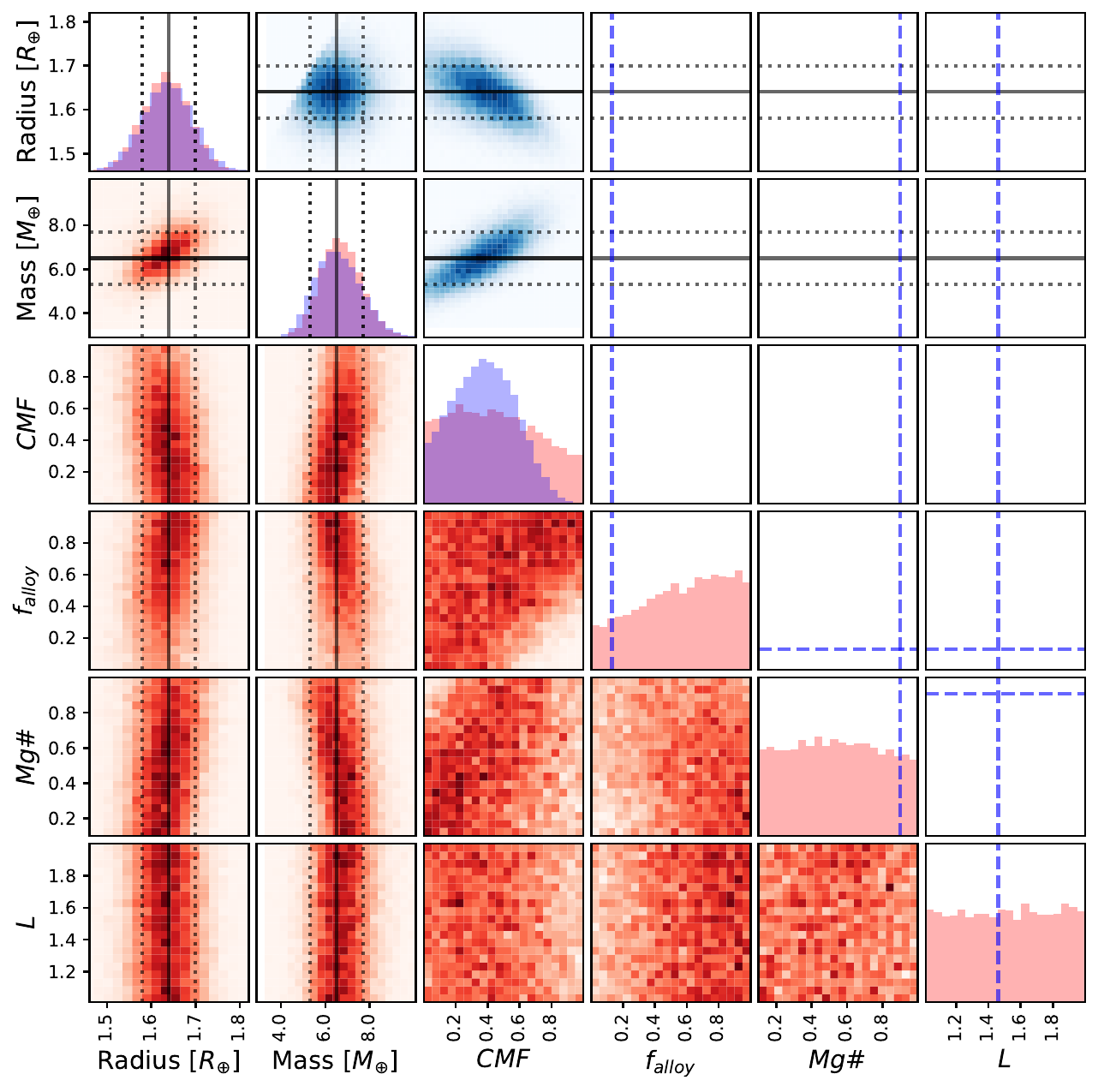}
    \caption{Comparison between posterior distributions of full and simple models for HD 15337 b. The blue and red density maps show the posterior distribution obtained by our retrievals of the simple model and the full model, respectively. The fixed parameters on the simple model are represented by blue dashed lines. The solid black lines represent the observed values of the mass and radius ($R_{p}$ and $M_{p}$) and the black dotted lines represent the observational uncertainties.  }
    \label{fig:fullvssimple}
\end{figure*}

Fig. \ref{fig:fullvssimple} compares the posterior distributions of the simple and full models. The CMF posterior in the full model exhibits a broader spread, as expected given the additional degeneracies introduced by $Mg\#$, $L$, and $f_{alloy}$. Nevertheless, the median value remains consistent with the simple model, falling within its one-sigma confidence interval. The full retrieval makes fewer assumptions about the chemical makeup of the planet and is therefore more realistic. For the rest of this work, we will use only the full retrievals.

\subsection{Devolatilization trends}

To study the devolatilization trend across our sample, we compute the normalized devolatilization ratio $Z_x$ for each planet, defined as the elemental molar fraction of element $x$ relative to Fe in the planet, normalized by the same ratio in the host star:

\begin{equation}
Z_{x} = \left( \frac{N_{x}}{N_{\mathrm{Fe}}} \right)_{p}
\Big/
\left( \frac{N_{x}}{N_{\mathrm{Fe}}} \right)_{\star}
\label{eq:Z}
\end{equation}

We use Fe as the reference element given its high condensation temperature (Table~\ref{tab:T50}) \citep{wang2019b}. The stellar mole ratio is computed from the observed abundances with Eq.~\ref{eq:rel_abun_star}:

\begin{equation}
\left( \frac{N_{x}}{N_{\mathrm{Fe}}} \right)_{\star}
=
\frac{10^{[x/H]} A_x}{10^{[\mathrm{Fe}/H]} A_{\mathrm{Fe}}}
\label{eq:rel_abun_star}
\end{equation}

where $A_x$ is derived from the photospheric solar abundance $\eta_x$ \citep{Asplund_2009} with Eq.~\ref{eq:A_x}:

\begin{equation}
A_x = 10^{(\eta_x - 12.00)}
\label{eq:A_x}
\end{equation}

The values of $T_{50\% , x}$ and $\eta_x$ used in this work are listed in Table~\ref{tab:T50} \citep{Lodders_2009,Asplund_2009}. We apply Eq.~\ref{eq:Z} to the posterior samples of the full retrieval and use a Gaussian kernel density estimator to obtain the median and $1\sigma$ uncertainties of $Z_x$. 

We then fit the devolatilization trend as a linear function of condensation temperature in logarithmic space (Eq.~\ref{eq:dev_trend}),

\begin{equation}
\log_{10}(Z_x) = m \, \log_{10}(T_{50\%,x}) + c
\label{eq:dev_trend}
\end{equation}

to the $Z_x$ values for each planet using MCMC with the \verb|emcee| package, obtaining the intercept and the best-fit slope, $m$, and its $1\sigma$ confidence interval. 

\begin{figure*}
\centering
\includegraphics[width=0.93\textwidth]{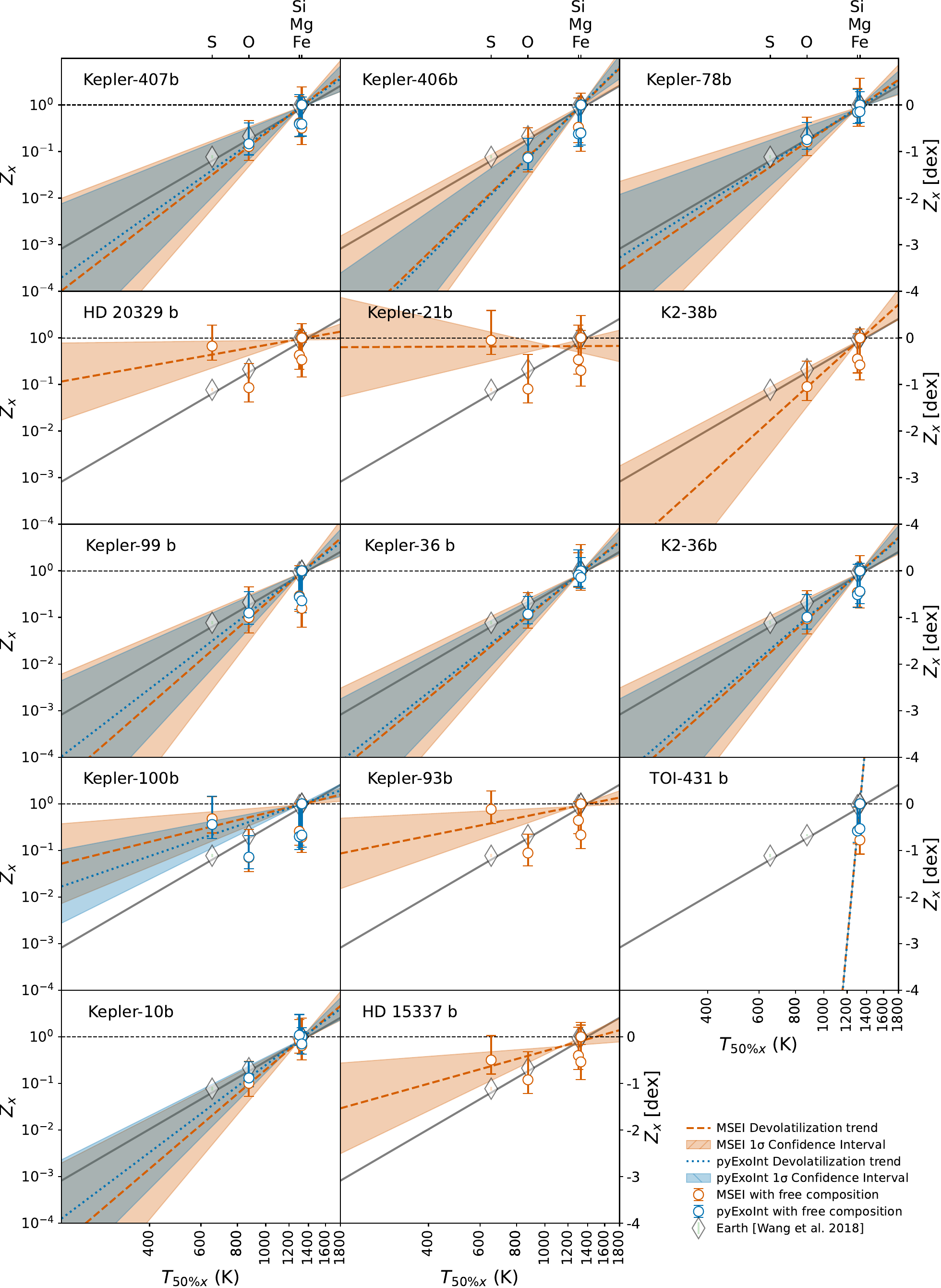}
\caption{Devolatilization trends from the full retrievals. Each panel shows the normalized devolatilization ratio ($Z_x$) plotted against the condensation temperature ($T_{50\%,x}$) of each element. The dashed (orange) and dotted (blue) lines indicate the devolatilization trends derived with MSEI and pyExoInt, respectively, while the solid gray line corresponds to the Earth's devolatilization trend. The shaded regions denote the $1\sigma$ confidence intervals around the retrieved trends.}
\label{fig:DT_full}
\end{figure*}

The resulting trends are shown in Fig.~\ref{fig:DT_full} and the fitted slopes for both forward models are listed in Table~\ref{tab:rel_mol_ratios}. MSEI and pyExoInt yield consistent slopes across all planets in the sample, with differences well within the uncertainties, demonstrating that the retrieved devolatilization trends are robust to the choice of interior structure models.

Of the 14 planets in our sample, 8 are compatible with an Earth-like devolatilization trend within the $1\sigma$ confidence interval, while 4 follow a trend with a lower slope than Earth. Kepler-21\,b shows a nearly flat devolatilization trend, suggesting little to no depletion of volatile elements relative to refractories. TOI-431\,b stands out as an outlier with an anomalously steep trend. This is likely an artifact of the few stellar abundances available for its host star, which yield a poorly constrained fit, and therefore we exclude it from further analysis.

\begin{table}[h]
\centering
\caption{Condensation temperatures and solar photospheric abundances for the elements used in the devolatilization trend fits.}
\label{tab:T50}
\begin{tabular}{lcc}
\hline
\hline
Element & $T_{50\%,x}$ (K) & $\eta_x$ \\
\hline
Fe & 1328 & $7.50 \pm 0.04$ \\
Mg & 1327 & $7.60 \pm 0.04$ \\
Si & 1302 & $7.51 \pm 0.03$ \\
O  & 875  & $8.69 \pm 0.05$ \\
S  & 655  & $7.12 \pm 0.03$ \\
\hline
\end{tabular}
\tablefoot{Condensation temperatures $T_{50\%,x}$ are from \cite{Lodders_2009}; solar photospheric abundances $\eta_x$ are from \cite{Asplund_2009}.}
\end{table}

\begin{table}[h]
\centering
\caption{Devolatilization trend slopes obtained from the full retrieval for both forward models.}
\label{tab:rel_mol_ratios}
\begin{tabular}{lcc}
\hline
\hline
Planet & MSEI slope [K$^{-1}$] & pyExoInt slope [K$^{-1}$] \\
\hline
Earth       & $3.65 \pm 0.03$ & --- \\
HD 15337 b  & $1.73 \pm 1.26$ & --- \\
HD 20329 b  & $1.12 \pm 1.04$ & --- \\
K2-36 b     & $5.61 \pm 2.61$ & $5.37 \pm 2.09$ \\
K2-38 b     & $5.70 \pm 2.39$ & --- \\
Kepler-100 b& $1.55 \pm 1.05$ & $2.16 \pm 0.97$ \\
Kepler-10 b & $5.37 \pm 2.12$ & $4.72 \pm 1.57$ \\
Kepler-21 b & $0.01 \pm 1.47$ & --- \\
Kepler-36 b & $5.03 \pm 2.15$ & $4.93 \pm 1.66$ \\
Kepler-406 b& $6.11 \pm 2.65$ & $6.23 \pm 1.87$ \\
Kepler-407 b& $4.89 \pm 2.39$ & $4.47 \pm 1.95$ \\
Kepler-78 b & $4.19 \pm 2.35$ & $3.94 \pm 1.68$ \\
Kepler-93 b & $1.25 \pm 0.94$ & --- \\
Kepler-99 b & $5.39 \pm 2.81$ & $4.81 \pm 2.00$ \\
TOI-431 b   & $67.22 \pm 0.00$ & $68.11 \pm 0.00$ \\
\hline
\end{tabular}
\tablefoot{Entries marked with --- indicate planets for which pyExoInt did not converge.}
\end{table}

\section{Discussion}
\label{sec:discussion}

\subsection{Comparison to previous work}

\begin{figure}
    \centering
    \includegraphics[width=0.8\linewidth]{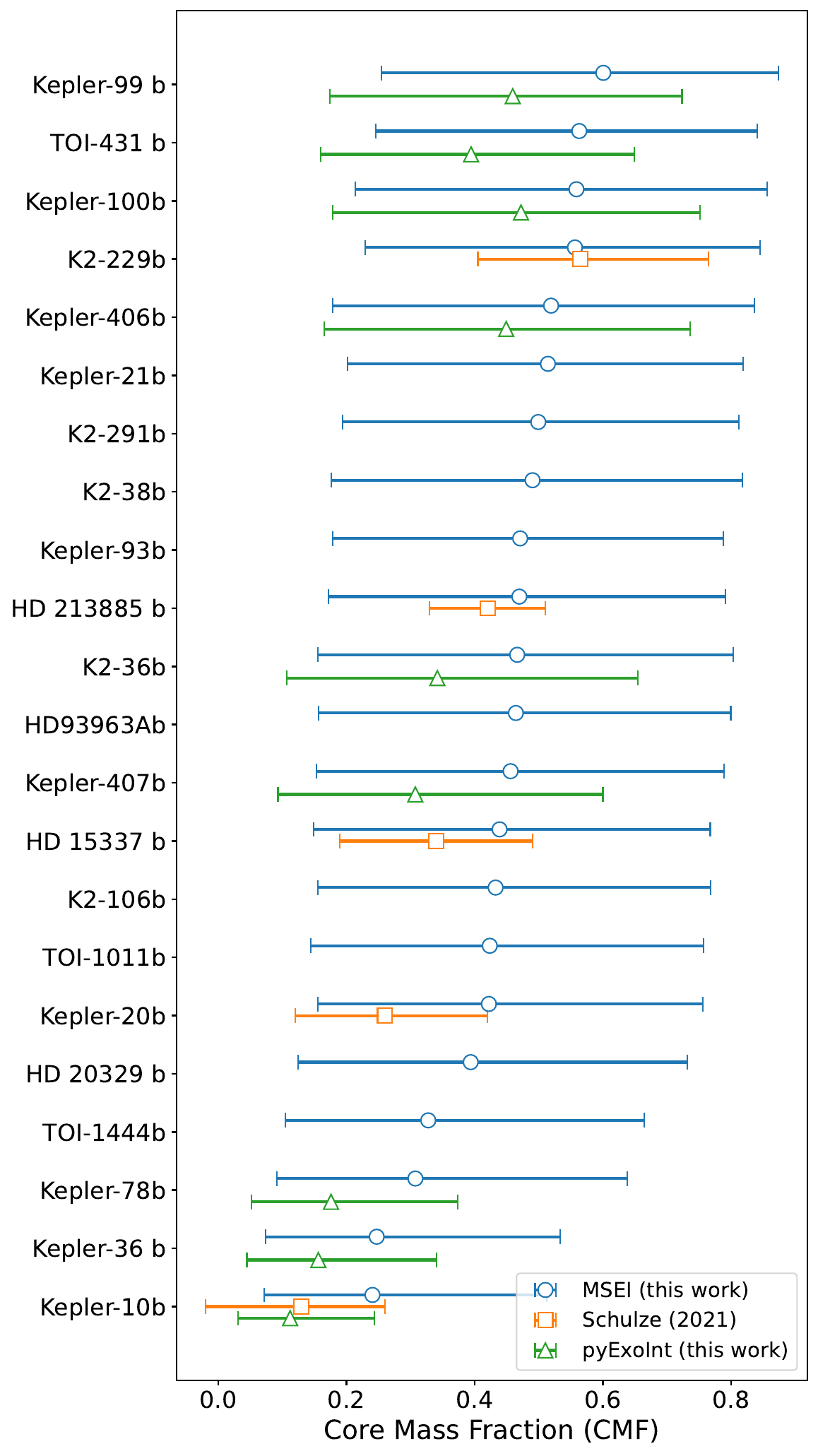}
    \caption{Comparison of core mass fraction (CMF) estimates from this work, obtained with MSEI (blue circles) and pyExoInt (green triangles), and from \cite{schulze_2021} (orange squares), where available. Error bars indicate $1\sigma$ uncertainties. Planets are ordered by ascending CMF from this work. Our CMF values are generally consistent with past work (see text).
}
    \label{fig:CMFschulze}
\end{figure}

In Fig.~\ref{fig:CMFschulze} we show a comparison of the retrieved CMFs using both MSEI and pyExoInt, as well as the values reported in \cite{schulze_2021}.
The CMFs derived from both pyExoInt and MSEI show good agreement with \cite{schulze_2021}, with estimates clustering around similar mean values. Our uncertainties are, however, larger, which we attribute to the more conservative nature of our approach: by leaving $f_{alloy}$, $\mathrm{Mg/Si}$ and $Mg\#$ as free parameters, our retrievals are subject to greater degeneracy, resulting in broader posterior distributions. In contrast, the CMF calculations in \cite{schulze_2021} fix the composition of the core to pure Fe, and the mantle is made of Fe-free silicates (MgSiO$_3$).  Kepler-10\,b stands out with a notably small CMF, consistent with the prediction of \cite{Wang2022}.

\subsection{Correlation between devolatilization trend slopes and planet properties}

Table \ref{tab:rel_mol_ratios} reveals a wide range of devolatilization slopes across our sample. To investigate the origin of this diversity, we compare the slopes against several system parameters that could explain them. \citet{Aguichine_2020} modeled rockline locations in protoplanetary disks as a function of heliocentric distance, finding that the relative iron mass fraction increases with distance and peaks in the vicinity of the rocklines. This suggests that the devolatilization slope may depend on orbital distance. We additionally examine planet mass and equilibrium temperature as potential influencing factors.

Neither pyExoInt nor MSEI retrieved abundances reveal a clear correlation between the devolatilization slope and orbital distance, equilibrium temperature, or planet mass (Fig.~\ref{fig:corr}). This absence of correlation may indicate that devolatilization is not primarily driven by these parameters. Alternatively, underlying trends may be hidden by the observational uncertainties in planetary masses, radii, and stellar abundances, which propagate into the compositional slopes.

Nevertheless, whether or not an individual planet exhibits a devolatilization trend carries significant implications for its formation history. The standard picture of devolatilization invokes nebular condensation, in which the thermal structure of the protoplanetary disk drives the selective removal of volatile elements. This produces a characteristic depletion pattern where elemental abundances scale with condensation temperature, consistent with solar system observations and the meteorite record \citep{Guimond_2024}. 
Conversely, the planets lacking a clear devolatilization signal may reflect distinct or more complex formation environments or evolutionary pathways, such as the pebble-accretion-based thermal processing paradigm \citep{Steinmeyer2023, Colmenares2024}.

Thermo-chemical equilibrium codes \citep[e.g., \texttt{GGchem},][]{Woitke_2018}, such as the open-source code \texttt{ECCOPLANETS} \citep{Timmermann_2023}, offer a theoretically motivated framework for constraining planetary compositions by simulating condensation sequences in the protoplanetary disk through Gibbs free energy minimization across a coupled system of gas and solid species. Given stellar abundances, disk pressure and temperature conditions, these codes predict the bulk composition of solids formed at each location in the disk. Applying them across a grid of disk parameters would allow us to simulate rocky-planet compositions and extract predicted devolatilization trends, directly comparable to those inferred from our MSEI and pyExoInt based retrievals. Such a comparison would identify the disk conditions that best reproduce the observed compositional signatures, thereby connecting measured planetary compositions to physically motivated formation environments.

\subsection{Systematic effects of Sulfur}

\begin{figure*}
    \sidecaption
    \includegraphics[width=12cm]{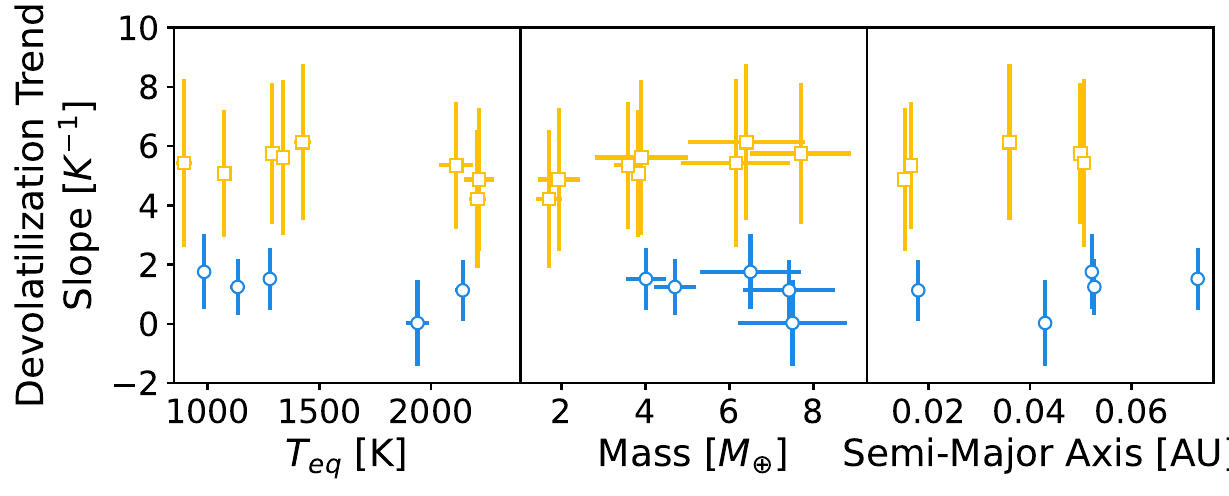}
    \caption{Correlation between the devolatilization trend slope calculated with MSEI and key planetary parameters: orbital distance, equilibrium temperature, and planet mass. The blue circles represent planets whose host stars have abundances available for all the elements in our forward model, including S. The orange squares represent planets whose host star lacks a measured S abundance. }
    \label{fig:corr}
\end{figure*}

As noted in Section \ref{sec:sample}, S abundance measurements are unavailable for a subset of the host stars in our sample. Planets orbiting stars without S measurements, and for which S is therefore excluded from the devolatilization fitting, tend to exhibit systematically steeper slopes compared to those with complete abundance measurements for Fe, Si, Mg, S, and O (Fig.~\ref{fig:corr}).

\subsection{Improving radius-measurement accuracy with \textit{PLATO}}

The PLAnetary Transits and Oscillations of stars (\textit{PLATO}) mission, a survey satellite scheduled for launch in 2026 \citep{plato}, offers a direct path toward addressing the observational limitations of the present work. By delivering more precise radius measurements for terrestrial planets around Sun-like stars at orbital distances beyond the current state of the art, \textit{PLATO} will improve CMF estimates and enable tighter constraints on planetary molar fractions. Furthermore, \textit{PLATO} will focus on rocky planets orbiting Sun-like stars, for which photospheric elemental abundances are more readily accessible than for late-type stars. This will substantially expand the sample of systems with both precise planetary parameters and well-characterized host star abundances, directly addressing one of the main limitations of the present work.

As discussed in Section~\ref{sec:sensitivity}, variations in $Mg\#$, $f_{alloy}$, and $\mathrm{Mg/Si}$ can induce radius changes of 8--11\%, 2--5\%, and $<$1\% across the two models, respectively. Constraining internal composition from the mass-radius relation alone therefore requires radius uncertainties smaller than the signal induced by each of these parameters. With an expected precision better than 3\% for Earth-sized planets, \textit{PLATO} will meet this threshold, allowing future retrievals to more robustly constrain compositional parameters and to determine devolatilization trends for individual planets with greater confidence. As demonstrated by \citet{Otegi_2020}, such improvements in radius precision, particularly when combined with accurate mass determinations from ground-based radial velocity follow-up, will substantially strengthen inferences of core mass fractions. 

\textit{PLATO} will further extend the accessible parameter space by characterizing rocky planets at orbital periods up to 365 days ($\sim$1~AU), and in favorable cases detecting planets on even wider orbits, including Mars analogs with periods of up to $\sim$680 days ($\sim$1.5~AU). This expanded baseline in orbital distance will enable a more robust investigation of how devolatilization varies with heliocentric distance.

\subsection{Determining composition with alternative methods}

Beyond improvements in transit observations, several complementary approaches exist for studying the elemental abundances of rocky exoplanets. Heavy elements accreted from planetary debris onto the white dwarf atmospheres serve as direct tracers of the bulk composition of the engulfed material \citep{Valencia_2025}. This pollution is thought to arise from the tidal disruption of minor bodies such as asteroids, dwarf planets, or fragments destabilized by surviving planets. These bodies form compact debris disks that gradually settle onto the stellar surface \citep{Jura_2003}. Spectroscopic analysis of these pollutants yields constraints on the bulk elemental makeup of the disrupted bodies \citep{Harrison_2018,Klein_2010,Zuckerman_2007}. Using UV spectroscopy of silicate features in the atmospheres of eight white dwarfs, \cite{rogers2025silicatemineralogybulkcomposition} inferred the mineralogy of the debris disks and showed that the silicate composition of planetary remnants is largely preserved from the original body. \cite{Xu__2019} found that most polluted white dwarfs exhibit elemental ratios consistent with those of Earth, though more volatile or water-rich compositions have also been reported, pointing to a diversity of progenitor materials \citep{Farihi_2013,Raddi_2015,Gentile_Fusillo_2017,Fans_nature_paper}. While interpretation can be complicated by measurement uncertainties and differential accretion of volatiles and refractories, polluted white dwarfs provide a rare direct window into exoplanetary building blocks.

A second complementary approach makes use of the tidal deformation of ultra-short-period planets. The strong gravitational interaction between such planets and their host stars distorts the shape of the planet from an approximate sphere into an ellipsoid, with the degree of deformation depending on mantle rigidity and hence on bulk composition. This shape change imprints a measurable signal on the transit light curve, modifying the inferred planetary radius in a composition-dependent way. \cite{Rogers_2020} incorporated tidal distortion corrections into the radius determination of KOI 1843.03 and found that accounting for the ellipsoidal shape significantly increases the inferred bulk density, exceeding values expected for Earth-like compositions. They further showed that the planet must contain at least 66\% iron in order to remain gravitationally bound against tidal disruption. The planetary shape thus provides an additional observational constraint on interior composition that is independent of the mass and radius, which can reduce the degeneracy of the interior retrievals.

\subsection{Outgassing and habitability}

Outgassing is a key process influencing the atmospheric and surface conditions of rocky exoplanets, and thus their potential habitability. Once the primordial atmosphere is stripped away by impacts and stellar irradiation, volatile species such as H$_2$O, CO$_2$, SO$_2$, and H$_2$ can be released from the mantle through volcanic and tectonic activity, contributing to the formation and long-term maintenance of secondary atmospheres \citep{Herbort_2020, Wordsworth_2022}. Thermochemical models and meteorite experiments suggest that early degassing alone can deliver several Earth ocean masses of H$_2$O to the surface \citep{Thompson_2021, Thompson_2023}. Sustained volcanism may continue replenishing atmospheric water over geological timescales, as has been proposed for planets in the TRAPPIST-1 system \citep{Thomas_2025}. Outgassing further regulates climate by supplying greenhouse gases such as CO$_2$ and H$_2$O, which raise surface temperatures and extend the range of stellar insolation compatible with liquid water.

The efficiency and character of outgassing are intimately tied to mantle composition and physical properties \citep{Spaargaren_2023}. The planetary molar fractions derived in this work can therefore inform the range of atmospheric types a given planet might plausibly host. The redox state of the mantle, for instance, controls volatile speciation: reduced interiors tend to produce CH$_4$- and H$_2$-dominated outgassing, while oxidized mantles favor CO$_2$-rich atmospheres \citep{Tian_2025}. Although we do not compute the redox state directly, the Fe/O ratio serves as a useful proxy, with higher oxygen availability per iron atom indicating a greater likelihood of iron residing in an oxidized state.

The $\mathrm{Mg/Si}$ ratio is another parameter sensitive to composition, as it governs mantle mineralogy and viscosity. Higher $\mathrm{Mg/Si}$ ratios promote olivine-dominated mantles with greater rigidity and viscosity, whereas lower ratios favor pyroxene-rich assemblages of lower viscosity \citep{Takeda_1998,Yamazaki_2001}. Lower-viscosity mantles drive more vigorous convection, which enhances volatile exchange between the deep interior and the surface, particularly during the magma-ocean stage, and facilitates the long-term replenishment of secondary atmospheres.

Mantle viscosity also regulates plate tectonics, which amplifies volatile exchange over geological timescales through subduction and arc volcanism. This continuous atmosphere-mantle coupling is critical for climate stability and habitability. On planets without plate tectonics, volcanic outgassing is the sole mechanism for building and sustaining an atmosphere \citep{Dorn_2018}, and its longevity depends on planetary mass, thermal state, and internal heating. Induction heating and tidal dissipation, for example, can sustain partial melting and episodic outgassing for billions of years, as proposed for Proxima b \citep{Noack_2021}.

Finally, the sulfur content of the mantle prior to core-mantle differentiation directly controls the flux of sulfur-bearing gases into the atmosphere. Elevated SO$_2$ and H$_2$S concentrations can be detrimental to habitability through acid rain, surface acidification, aerosol formation, and reduced solar insolation at the surface. In smaller quantities, however, these gases may support prebiotic chemistry and signal active volcanism which is itself a driver of volatile cycling and a potential energy source for life. Overall, their impact on habitability depends on concentration, duration, and atmospheric context.

\section{Conclusion}

We employed two independent interior structure models, MSEI and pyExoInt, combined with MCMC methods, to retrieve the interior compositions of a sample of rocky exoplanets with measured host stellar abundances. Using a precomputed grid of interior models to accelerate the retrieval, both forward models estimate the molar fractions of Fe, Si, Mg, S, and O from observed planetary masses and radii. This enables a quantitative assessment of whether devolatilization can account for the iron enrichment observed in these rocky exoplanets. A comparison between MSEI and pyExoInt shows consistent results across retrieved quantities.

The forward models suggest that for a planet with mass and CMF values of the Earth, $Mg\#$ has the largest impact on predicted planetary radius ($\sim$8\% for MSEI and $\sim$11\% for pyExoInt), followed by $f_{alloy}$ ($\sim$5\% for MSEI and $\sim$2\% for pyExoInt), while $\mathrm{Mg/Si}$ remains below the sensitivity threshold of current observations ($<$1\% for both models). The precomputed grids of both forward models are publicly available on Zenodo\footnote{\url{https://doi.org/10.5281/zenodo.24065474}}.

Eight out of fourteen planets exhibit an Earth-like devolatilization trend within $1\sigma$, four show a less-devolatilized trend, and Kepler-21 b displays a near-flat trend consistent with a stellar-like bulk composition. We found no significant correlation between the devolatilization slope and planetary mass, equilibrium temperature, or semi-major axis, likely reflecting the complexity of the planet formation process and/or the limited sample size and yet-large observational uncertainties.

Our results are compatible with the view that rocky exoplanets commonly undergo volatile depletion processes analogous to those inferred for Earth, and that such devolatilization can contribute to the apparent iron enrichment inferred from their bulk densities. We stress, however, that we do not claim devolatilization alone accounts for the full range of observed compositions; other mechanisms such as collisional stripping \citep{Scora_2020, scora2022} and Fe nucleation \citep{johansen_dorn2022} may still be required for the most iron-rich planets. Our results also reveal significant compositional diversity, which has direct implications for habitability, as the interior volatile inventory governs secondary atmosphere composition through outgassing. The \textit{PLATO} mission, with its expected radius precision below 3\% for Earth-sized planets and access to longer orbital periods, will enable more precise retrievals and allow us to test whether devolatilization trends vary with formation location in the disk. To further validate or constrain our modeling results, it would also be extremely valuable to determine an accurate and homogeneous set of elemental abundances of the host stars of \textit{PLATO} rocky exoplanet targets.

This work demonstrates that except for highly refractory elements (e.g. Ca, Al, and Ti) stellar composition cannot serve as a direct proxy for planetary composition: devolatilization could significantly alter key elemental ratios (including Fe/Si), producing compositional offsets between a star and its rocky planets. Accounting for these processes is essential when interpreting exoplanetary compositions and assessing habitability, and underscores the need for improved observational constraints and larger, homogeneous planet samples.

\section*{Data availability}
The precomputed grids of interior structure models for MSEI and pyExoInt are publicly available on Zenodo at \url{https://doi.org/10.5281/zenodo.24065474}.

\begin{acknowledgements}

We thank the anonymous referee for their constructive comments, which have improved the quality of the manuscript. The research shown here acknowledges use of the Hypatia Catalog, an online compilation of stellar abundance data as described in \cite{hypatia}. This research has made use of the NASA Exoplanet Archive, which is operated by the California Institute of Technology, under contract with the National Aeronautics and Space Administration under the Exoplanet Exploration Program. L. A-A. thanks Bastien Brugger, Magali Deleuil and Olivier Mousis for the development of the Marseille Interior Structure model, which was used to obtain the results presented in this paper. L. A-A. thanks Caroline Dorn for useful discussion on the role of sulfur in rocky planet interiors. H.S.W. acknowledges support from the Carlsberg Foundation through the FIRSTATMO project. 
\end{acknowledgements}

\bibliographystyle{bibtex/aa} 
\bibliography{refs}

\begin{appendix}

\section{Equilibrium temperature calculations}
\label{sec:equilibrium_temp_appendix}

The equilibrium temperature of a planet is given by
\begin{equation}
    T_{eq} = T_{\text{eff}} \left( \frac{R_{\star}}{2a} \right)^{\frac{1}{2}} (1 - A_b)^{\frac{1}{4}},
    \label{eq:planet_equilibrium_temp}
\end{equation}
where $T_{\text{eff}}$ and $R_{\star}$ are the effective temperature and radius of the host star, $a$ is the semi-major axis, and $A_b$ is the Bond albedo, which we set to zero, as in previous work such as \citet{Adibekyan_2021}. The semi-major axis follows from the orbital period $\mathcal{T}$ through Kepler's third law,
\begin{equation}
    \mathcal{T}^2 = \frac{4\pi^2}{G (M_{\star} + M_p)} a^3,
    \label{eq:kepler_third}
\end{equation}
where $M_{\star}$ is the stellar mass, $M_p \ll M_{\star}$ the planet mass, and $G$ the gravitational constant. Uncertainties on $T_{eq}$ are propagated by Monte Carlo sampling of $T_{\text{eff}}$, $R_{\star}$, and $M_{\star}$; the contribution of the period uncertainty is negligible.

\section{Extra tables}

\begin{sidewaystable*}
\centering
\caption{Retrieved mole fractions of rocky exoplanets from the MSEI and pyExoInt interior models.}
\label{tab:retrieved_abundances}
\setlength{\tabcolsep}{4pt}
\begin{tabular}{lccccc|ccccc}
\hline
\hline
 & \multicolumn{5}{c}{MSEI} & \multicolumn{5}{c}{pyExoInt} \\
\hline
Planet & Fe & Si & Mg & S & O & Fe & Si & Mg & S & O \\
\hline
K2-36 b & $1.002^{+0.205}_{-0.173}$ & $0.346^{+0.478}_{-0.250}$ & $0.355^{+0.636}_{-0.281}$ & --- & $0.093^{+0.131}_{-0.069}$ & $0.996^{+0.207}_{-0.171}$ & $0.308^{+0.462}_{-0.195}$ & $0.363^{+0.501}_{-0.221}$ & --- & $0.101^{+0.113}_{-0.059}$ \\
Kepler-407 b & $0.998^{+0.201}_{-0.168}$ & $0.409^{+0.537}_{-0.279}$ & $0.318^{+0.647}_{-0.256}$ & --- & $0.128^{+0.165}_{-0.086}$ & $1.001^{+0.201}_{-0.173}$ & $0.391^{+0.571}_{-0.243}$ & $0.388^{+0.519}_{-0.231}$ & --- & $0.146^{+0.150}_{-0.080}$ \\
Kepler-78 b & $1.001^{+0.204}_{-0.171}$ & $0.684^{+0.825}_{-0.467}$ & $0.727^{+1.186}_{-0.532}$ & --- & $0.161^{+0.198}_{-0.109}$ & $0.997^{+0.202}_{-0.169}$ & $0.697^{+0.785}_{-0.391}$ & $0.716^{+0.707}_{-0.386}$ & --- & $0.180^{+0.155}_{-0.086}$ \\
TOI-431 b & $0.999^{+0.096}_{-0.088}$ & $0.266^{+0.304}_{-0.175}$ & $0.167^{+0.266}_{-0.117}$ & --- & --- & $1.001^{+0.100}_{-0.088}$ & $0.262^{+0.312}_{-0.149}$ & $0.293^{+0.308}_{-0.160}$ & --- & --- \\
Kepler-36 b & $0.999^{+0.199}_{-0.168}$ & $0.829^{+0.901}_{-0.518}$ & $0.807^{+1.217}_{-0.598}$ & --- & $0.112^{+0.124}_{-0.073}$ & $0.998^{+0.200}_{-0.171}$ & $0.813^{+1.010}_{-0.454}$ & $0.720^{+0.705}_{-0.383}$ & --- & $0.118^{+0.102}_{-0.055}$ \\
K2-38 b & $0.999^{+0.205}_{-0.169}$ & $0.353^{+0.441}_{-0.243}$ & $0.262^{+0.465}_{-0.194}$ & --- & $0.090^{+0.114}_{-0.061}$ & --- & --- & --- & --- & --- \\
Kepler-100 b & $0.999^{+0.101}_{-0.087}$ & $0.258^{+0.351}_{-0.179}$ & $0.194^{+0.359}_{-0.149}$ & $0.479^{+0.525}_{-0.336}$ & $0.070^{+0.098}_{-0.050}$ & $1.003^{+0.094}_{-0.089}$ & $0.198^{+0.297}_{-0.124}$ & $0.213^{+0.290}_{-0.127}$ & $0.360^{+0.504}_{-0.252}$ & $0.072^{+0.076}_{-0.042}$ \\
Kepler-99 b & $1.001^{+0.203}_{-0.170}$ & $0.290^{+0.435}_{-0.224}$ & $0.156^{+0.294}_{-0.144}$ & --- & $0.100^{+0.152}_{-0.076}$ & $1.003^{+0.204}_{-0.170}$ & $0.276^{+0.390}_{-0.171}$ & $0.228^{+0.287}_{-0.134}$ & --- & $0.125^{+0.132}_{-0.070}$ \\
HD 15337 b & $1.001^{+0.579}_{-0.377}$ & $0.395^{+0.555}_{-0.271}$ & $0.289^{+0.563}_{-0.251}$ & $0.317^{+0.382}_{-0.219}$ & $0.119^{+0.166}_{-0.082}$ & --- & --- & --- & --- & --- \\
HD 20329 b & $0.997^{+0.260}_{-0.209}$ & $0.426^{+0.517}_{-0.296}$ & $0.338^{+0.605}_{-0.284}$ & $0.667^{+0.685}_{-0.469}$ & $0.085^{+0.102}_{-0.061}$ & --- & --- & --- & --- & --- \\
Kepler-10 b & $0.998^{+0.261}_{-0.207}$ & $0.843^{+0.868}_{-0.542}$ & $0.646^{+0.873}_{-0.454}$ & --- & $0.100^{+0.108}_{-0.064}$ & $1.001^{+0.262}_{-0.211}$ & $1.083^{+1.106}_{-0.539}$ & $0.704^{+0.570}_{-0.353}$ & --- & $0.131^{+0.103}_{-0.057}$ \\
Kepler-21 b & $1.008^{+1.100}_{-0.544}$ & $0.340^{+0.579}_{-0.245}$ & $0.199^{+0.396}_{-0.154}$ & $0.889^{+1.309}_{-0.614}$ & $0.080^{+0.137}_{-0.056}$ & --- & --- & --- & --- & --- \\
Kepler-93 b & $1.000^{+0.200}_{-0.170}$ & $0.441^{+0.412}_{-0.286}$ & $0.215^{+0.307}_{-0.146}$ & $0.761^{+0.700}_{-0.536}$ & $0.088^{+0.081}_{-0.057}$ & --- & --- & --- & --- & --- \\
Kepler-406 b & $1.000^{+0.097}_{-0.089}$ & $0.332^{+0.482}_{-0.255}$ & $0.253^{+0.495}_{-0.229}$ & --- & $0.078^{+0.111}_{-0.059}$ & $1.001^{+0.098}_{-0.093}$ & $0.239^{+0.319}_{-0.149}$ & $0.249^{+0.310}_{-0.146}$ & --- & $0.073^{+0.071}_{-0.042}$ \\
\hline
\end{tabular}
\tablefoot{Values are the posterior median with asymmetric $1\sigma$ uncertainties, normalized relative to Fe for both models. Entries marked with --- indicate that the element was unconstrained or absent from the model fit.}
\end{sidewaystable*}

\end{appendix}
\end{document}